# Spin-Transfer Effects in Nanoscale Magnetic Tunnel Junctions

G. D. Fuchs, N. C. Emley, I. N. Krivorotov, P. M. Braganca, E. M. Ryan, S. I. Kiselev, J. C. Sankey, D. C. Ralph and R. A. Buhrman

Cornell University, Ithaca NY 14853-2501

J. A. Katine

Hitachi Global Storage Technologies, San Jose CA 95120

## Abstract

We report measurements of magnetic switching and steady-state magnetic precession driven by spin-polarized currents in nanoscale magnetic tunnel junctions with low-resistance, $< 5\ \Omega{-}\mu m^2$, barriers. The current densities required for magnetic switching are similar to values for all-metallic spin-valve devices. In the tunnel junctions, spin-transfer-driven switching can occur at voltages that are high enough to quench the tunnel magnetoresistance, demonstrating that the current remains spin-polarized at these voltages.

The transfer of spin angular momentum from a spin-polarized current to a nanoscale ferromagnet (F) can reversibly switch the orientation of the magnet's moment[1] or can excite the magnet into microwave-frequency precession[2]. This spin-transfer effect has the potential to provide a scalable means of writing information in magnetic random access memories (MRAM)[3,4] that would be non-volatile, dense and high speed. Spin-transfer switching, however, has been fully demonstrated only in all-metallic systems fabricated from F / normal metal / F multilayers. Such samples have low resistances (~ 10 Ω), much less than optimal values (>1 kΩ) for MRAM circuits. One approach to reach higher resistance is to use magnetic tunnel junction (MTJ) devices. This poses considerable challenges, since to achieve spin-transfer switching the devices must be able to withstand high current densities (~$10^7$ A/cm$^2$) without exceeding the breakdown voltage of the barrier[5].

Here we discuss measurements of magnetic switching and spin-transfer-driven precession in magnetic tunnel junctions with ultra-thin barriers, having specific resistances *(RA)* in the range 1–4 Ω–μm$^2$. We observe reproducible spin-transfer switching between parallel and anti-parallel magnetic alignments of the electrodes, as revealed by changes in the zero-bias device resistance Δ*R*. Δ*R* as large as 160 Ω have been achieved with a MTJ having a resistance of 1760 Ω in the anti-parallel state. With different MTJ devices we have also been able to excite steady-state precessional magnetic modes using DC currents, similar to dynamical behavior observed in metal spin valves[2].

Our devices were prepared by first sputtering a thin-film multilayer onto a thermally oxidized silicon wafer. The layers for our first set of devices consisted of: Py 4/Cu 80/Ta 10/CoFeB 8/Al 0.65(Ox) /CoFeB 2/Cu 5/Pt 30 (thickness in nm). Here Py is $Ni_{81}Fe_{19}$, CoFeB is $Co_{88.2}Fe_{9.8}B_2$, and Ox represents a 300 K thermal oxidation of the aluminum layer (27 torr-s). E-beam lithography, photolithography, and ion milling[1] were used to pattern the layers into elliptical nanopillars to provide shape anisotropy. Both the thinner ("free") and thicker ("fixed") magnetic electrodes were patterned as illustrated in Fig. 1a, so that the free layer experiences a dipole field $H_d$ from the fixed layer.

Figure 1a shows the 300 K tunneling magnetoresistance (TMR) of a MTJ with size ~ 40 × 130 $nm^2$. The TMR of 7% is in the range observed previously for very thin $AlO_x$ tunnel barriers with low specific resistance[5], in this instance 3.3 $\Omega-\mu m^2$. The reduction of the TMR from the larger values that can be obtained with thicker barriers can be attributed, at least in part, to the presence of leakage channels penetrating the ultra-thin barrier. The free layers in this sample is a single, super-paramagnetic particle. At room temperature the single domain free layer, due to its small area and thickness, switches non-hysteretically due to thermal activation[6] when the device is cycled through its minor loop, *i.e.* when an external field $H$ is applied along the major axis of the nanomagnet and varied about the value $H = H_d = 325$ Oe (Fig.1b). Near this bias field, a small DC current can be used to manipulate the free layer reversibly between orientations parallel (P) and anti-parallel (AP) to the fixed layer. In Fig. 1c we display the differential resistance *dV/dI vs. I* at three value values of $H$. For $H = 307$ Oe, the two layers remain AP, while for $H = 408$ Oe they are P. When $H = 367$ Oe, the free layer moves between the P and AP alignment as *I* is ramped from 0 to 0.1 mA. This trace is noisy because the free layer exhibits telegraph-type switching between the P and AP states in this current range. We note that the area of this MTJ device is sufficiently small that current-induced vortex states do not occur[7].

When cooled to 77 K, the thermal excitations are reduced and the device exhibits magnetic hysteresis (Figs. 2a and 2b)[6]. In the minor magnetoresistance loop (Fig. 2b), the transition of the free layer from P to AP alignment is not abrupt but extends over a range of $H$. We tentatively attribute this behavior to Néel coupling between the magnetic electrodes[8] that can stabilize intermediate magnetic orientations.

At a fixed $H$, the current can be used to switch the device reproducibly between the P and AP orientations at 77 K (Fig. 2(c)). The resistance change at zero bias, $\Delta R = 40\ \Omega$, is equal to the change between P and AP states driven by $H$. The switching from AP to P alignment at $I = -0.46$ mA is fairly abrupt, while the transition from P to AP occurs gradually near $I = 0.58$ mA, behavior that mirrors the field sweeps in Fig. 2b.

In Fig. 2d we show the magnetoresistance minor loop and the current-resistance loop at 77 K for a different, higher-resistance MTJ of smaller area ($\sim 25 \times 112$ nm$^2$). This device had $\Delta R = 180\ \Omega$, TMR = 11%, and $RA = 3.5\ \Omega-\mu m^2$ at 77K. We attribute the higher TMR and $RA$ to the presence of fewer shunting channels through the barrier. By applying bias loops repeatedly to higher maximum values, we determined the critical biases for current-induced switching in both bias directions (indicated by arrows). In general, it is known that MTJs exhibit a gradual decrease of TMR with increasing voltage *(V)*, the rate of which depends on barrier quality, thickness and material[9,10]. In Fig. 2d, current-induced switching occurs at a sufficiently large $V$ that the TMR is effectively quenched, so that the switching is not apparent until $V$ is reduced to lower values. Since a current must maintain a spin polarization in order to provide spin-transfer torque, this observation demonstrates that the absence of TMR at high bias does not represent the absence of spin polarization in $I$, but rather the lack of a dependence of junction resistance to that polarization.

The critical current density $J_c$ for spin-transfer switching of a nanomagnet in the absence of a net external magnetic field[1,11,12] is predicted to be $J_c = \alpha e M t [H_K + 2\pi M]/h\eta$ where $M$ is the magnetization, $t$ the thickness, $H_K$ the anisotropy field, $\alpha$ the Gilbert damping parameter, and $\eta$ an efficiency factor. This does not include thermal fluctuation effects which act to reduce $J_c$ from the zero-T value[6,12,13]. We find it interesting to compare the values of $J_c$ we measure for CoFeB free layers in MTJs to previous measurements of Co layers in metal spin valves[12], since the magnetization ($M$) of CoFeB, (1180 emu/cm$^3$) is comparable to Co (1400 emu/cm$^3$). We note that the Co devices studied in ref. [12] were larger than the MTJs discussed here, and they were measured at room temperature rather than 77 K, but we do not expect these differences to alter $J_c$ by more than a factor of 2. For Co spin-valve devices the average value of $J_c$ was $1.6 \times 10^7$ A/cm$^2 \times t$/nm. In the MTJ samples we find $J_c = (0.5\text{-}0.8) \times 10^7$ A/cm$^2 \times t$/nm. We have considered whether self-heating by high bias levels might reduce $J_c$ in the MTJs. We will discuss the temperature dependence of $J_c$ elsewhere, but the strong changes that we see in cooling from 300 K to 77 K indicate that self-heating is not a dominant effect. Therefore we conclude that the spin-transfer efficiency of the current in the MTJs is at least comparable to that in spin valve nanopillars.

We have also observed at room temperature the excitation of steady-state magnetic precession induced by spin transfer, in MTJs that were more heavily shunted, most likely in part by sidewall re-deposition during the ion-milling process. These devices therefore had lower values of $R$ (65 Ω), $RA$ (~ 1 Ω–μm$^2$), and TMR (2.3 %). The free layer was 2-nm-thick Py, patterned into an ellipse ~ 95 × 195 nm$^2$. The counter electrode was a Py/Co bilayer coupled to an antiferromagnetic IrMn layer. When the device was biased at a value of $H$ to set the layers anti-parallel (Fig. 3a), negative current (electrons flowing from the fixed to the free layer) excited the free layer into oscillation in the 1–5 GHz frequency range, as detected by the high frequency TMR signal[2]. Positive $I$ produced no oscillations. When $H$ was used to set the moments in the parallel configuration (Fig. 3b), positive $I$ excited oscillations, while negative $I$ did not. The oscillation frequency varied with $H$ in good agreement with the Kittel formula for ferromagnetic resonance[14]. These results are in accord with predictions for the spin-transfer effect[11] and are similar to observations in metallic spin valves[2]. Therefore, shunted MTJs have potential for high impedance spin-transfer-driven oscillator applications.

In summary, we have observed reproducible switching and microwave oscillations of nanomagnets driven by spin-transfer from currents passing between a fixed ferromagnet layer and the nanomagnet via tunnel barriers of low specific resistance $< 5 \, \Omega{-}\mu m^2$. Despite values of TMR that are lower than for MTJs with more resistive barriers, the critical currents for spin-transfer switching in these junctions are comparable to values found for metallic spin-valve nanopillars. This indicates that spin transport via these thin and imperfect barriers is a robust and efficient process. Spin-transfer switching can occur at bias values where the TMR of the junction is zero, demonstrating that the absence of TMR at high bias does not mean that the junction current is un-polarized. These findings are encouraging for the prospect of practical MTJ spin-transfer-switched MRAM and nano-oscillator device technology.

We acknowledge support from the Army Research Office, from the NSF/NSEC program through the Cornell Center for Nanoscale Systems, and from DARPA through Motorola. We also acknowledge use of the NSF-supported Cornell Nanofabrication Facility/NNIN and the facilities of the Cornell Center for Materials Research.

# Figure Captions

**Fig. 1:** a) TMR major loop at room temperature for device 1. b) TMR minor loop for the free layer of device 1 at room temperature, with the fixed layer aligned with $H$. The free layer is super-paramagnetic, so there is no hysteresis. c) Current loops at room temperature for device 1. At 367 Oe, an applied current can shift the free layer between the parallel (408 Oe) and antiparallel (307 Oe) orientations.

**Fig. 2:** a) TMR major loop at 77 K for device 1. b) TMR minor loop for the free layer of device 1 at 77 K, with the fixed layer aligned with $H$. c) Current-resistance loop at 77 K for device 1, with $H \sim H_d$. d) Current-resistance loop at 77 K for sample 2, with $H \sim H_d$. Arrows at -0.40 mA and 0.38 mA mark the switching currents. Inset: TMR minor loop for the free layer of device 2 at 77 K, with the fixed layer aligned with $H$.

**Fig. 3:** Spectra of microwave signals generated by DC currents at fixed $H$ for device 3 when a) the two magnetic electrodes have been oriented anti-parallel at $I=0$ and b) the two layers have been oriented parallel at $I=0$.

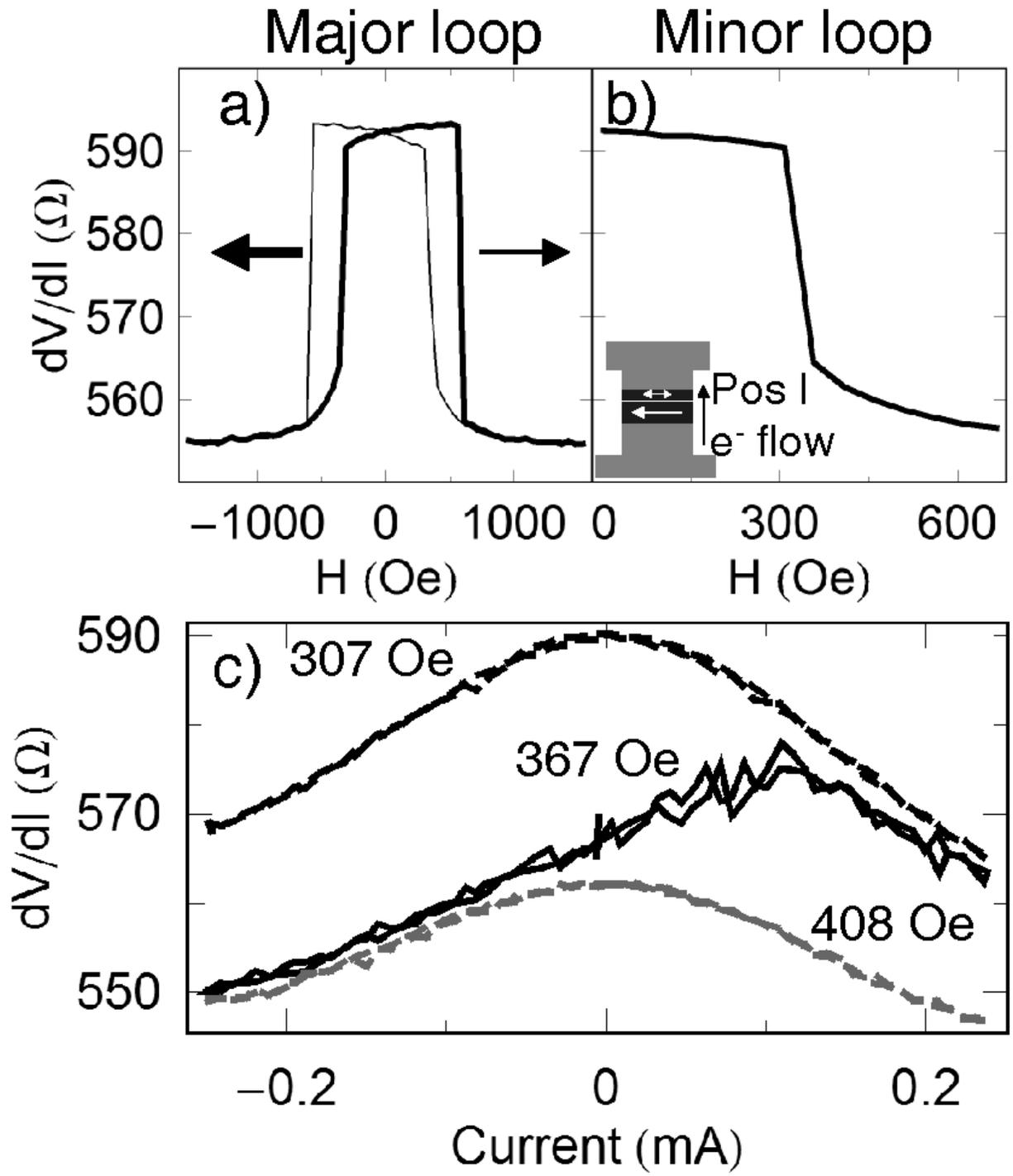

Fuchs *et al.* Figure 1

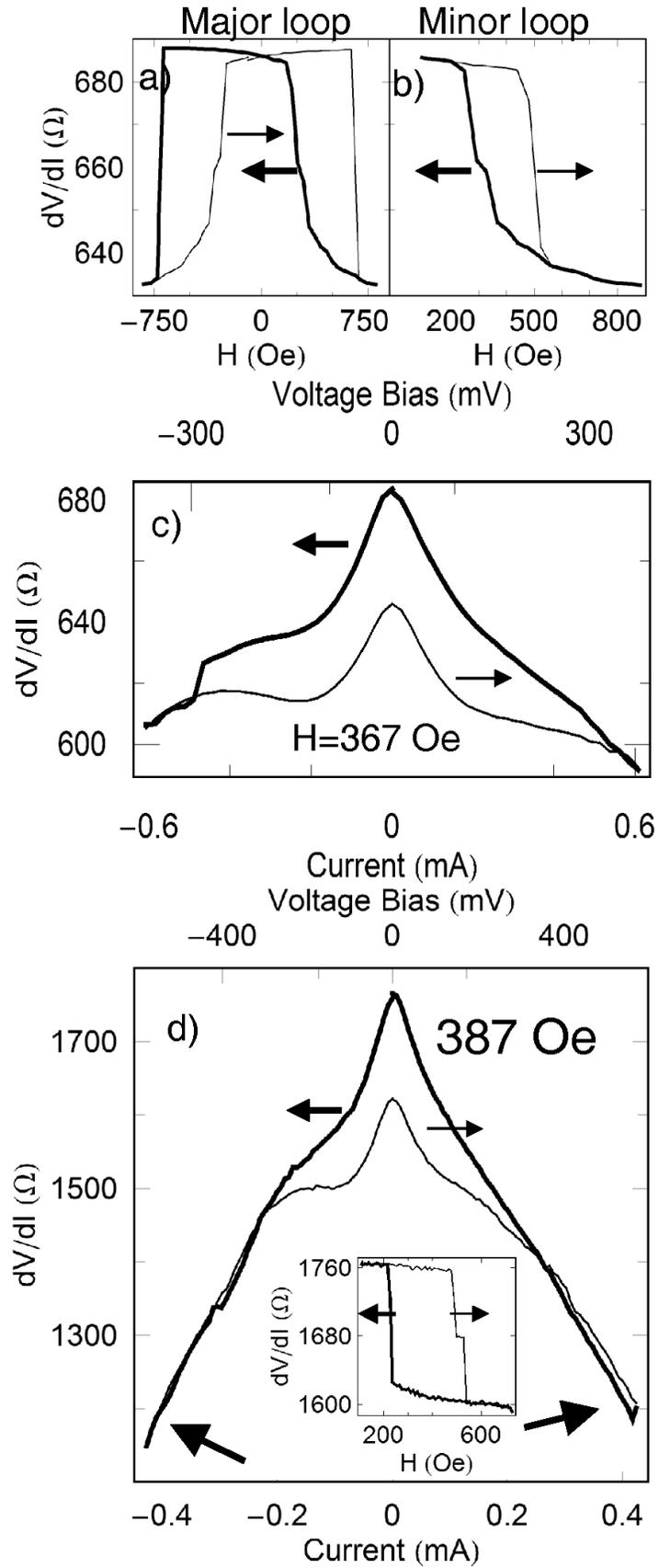

Fuchs *et al*. Fig 2

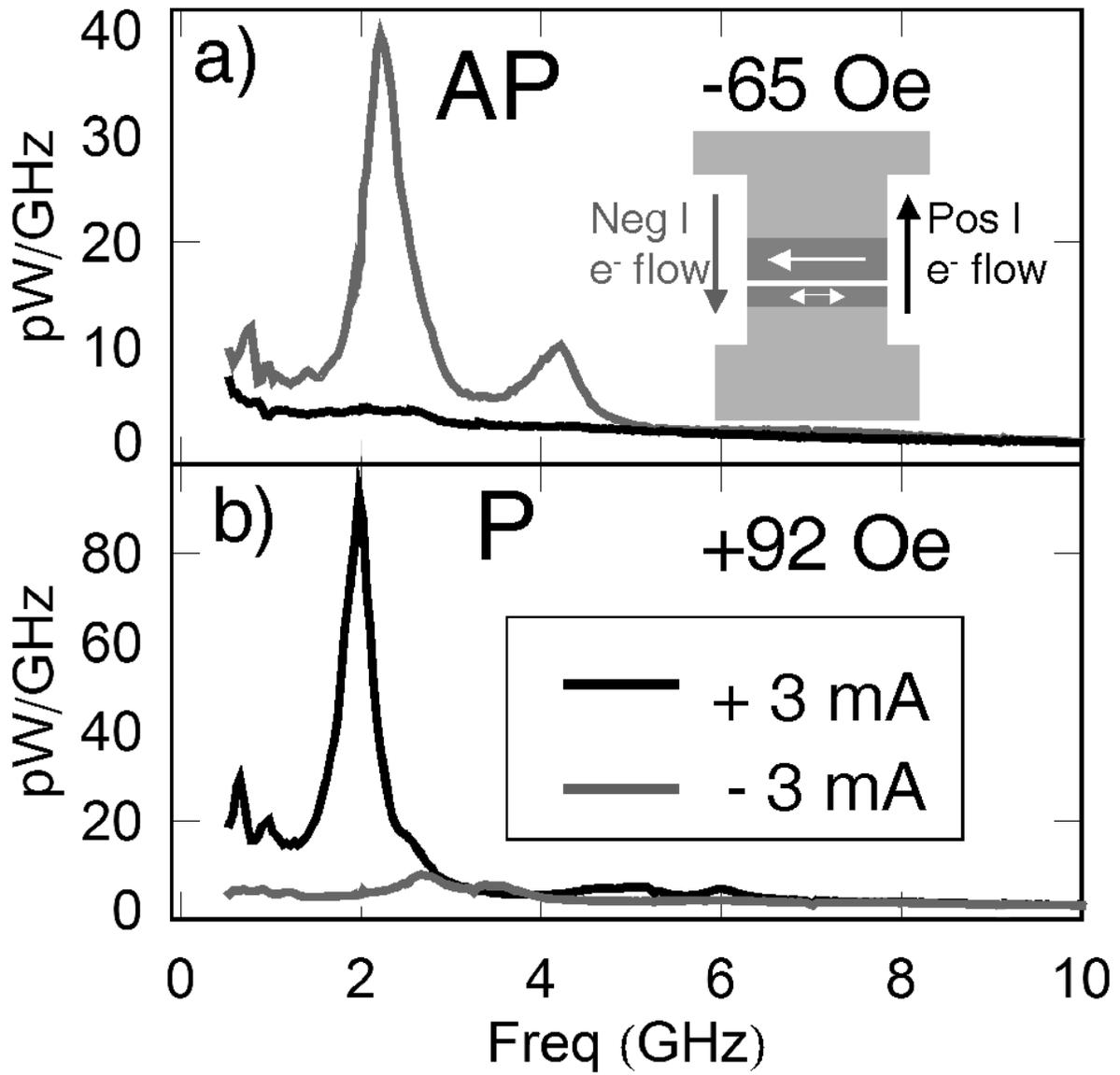

Fuchs *et al.* Figure 3

12